\documentstyle[12pt]{article}

\def\R{I\kern-.30em{R}}
\def\N{I\kern-.30em{N}}
\def\F{I\kern-.30em{F}}
\def\C{I\kern-.60em{C}}
\def\Z{Z\kern-.50em{Z}}
\def\T{I\kern-.30em{T}}

\newcommand{\rf}[1]{(\ref{#1})}

\def\szj{{S^{z}\!}_j}

\def\fju{{f}_{j'\uparrow}}
\def\fjd{{f}_{j'\downarrow}}

\def\diu{{d}_{i\uparrow}}
\def\did{{d}_{i\downarrow}}

\def\diuc{{d^\dagger}_{{i}\uparrow}}
\def\didc{{d^\dagger}_{{i}\downarrow}}

\def\fjuc{{f^\dagger}_{{j}\uparrow}}
\def\fjdc{{f^\dagger}_{{j}\downarrow}}

\newcommand{\beq}{ \begin{equation}}
\newcommand{\eeq}{\end{equation}}
\newcommand{\ba}{\begin{array}}
\newcommand{\ea}{\end{array}}
\newcommand{\beqa}{\begin{eqnarray}}
\newcommand{\eeqa}{\end{eqnarray}}

\font\twelve=cmbx10 at 15pt
\font\ten=cmbx10 at 12pt

\begin{document}

\tolerance=2000

\begin{titlepage}

\begin{center}

\renewcommand{\thefootnote}{\fnsymbol{footnote}}

{\ten Centre de Physique Th\'eorique\footnote{
Unit\'e Propre de Recherche 7061} - CNRS - Luminy, Case 907}
{\ten F-13288 Marseille Cedex 9 - France }

\vspace{1 cm}

{\twelve MOTION OF A SINGLE HOLE \\
IN A DISORDERED MAGNETIC BACKGROUND}

\vspace{0.3 cm}

\setcounter{footnote}{0}
\renewcommand{\thefootnote}{\arabic{footnote}}

{\bf A. BELKASRI\footnote{and Universit\'e d'Aix-Marseille II}
 and J.L. RICHARD}

\vspace{2 cm}

{\bf Abstract}

\end{center}

The spectrum of a single hole is calculated within the spin-hole
model\break
using a variational method. This calculation is done for
any rotational invariant magnetic background. We have found that when
the magnetic  background  changes from a disordered to a locally
ordered state, the spectrum changes qualitatively. We have also found
that the spin pattern around the hole is polarized. This problemis
related to the study of copper oxide planes $CuO_2$ doped with a small
number of holes.

\vspace{2 cm}


\bigskip

\noindent Number of figures : 3

\bigskip

\noindent July 1994

\noindent CPT-94/P.3058

\bigskip

\noindent anonymous ftp or gopher: cpt.univ-mrs.fr

\end{titlepage}

\section{Introduction}
	The study of the motion of one  hole in strongly correlated systems
is very important  because, in principle, it may contain some features
of the properties of the  $CuO_2$ planes doped with a small number of
holes. The one hole motion in a quantum  antiferromagnetic (AFM)
background has been studied by
several authors\cite{Varma,Kane,MHorsch,JLR} in the framework of the
$t$-$J$ model. These studies were carried out within  an AFM ordered
state. The main result of these analysis is the existence of a
well-defined quasiparticle description for a coherent hole propagation
with the band minima at $(\pm\pi/2, \pm\pi/2)$ and the bandwidth $W$
of the order of $J$ and not $t$.  These results are in good agreement
with numerical exact diagonalization for small
systems\cite{PoilSchulz}\cite{SHorsch}.  All these investgations
concern the magnetic ordered case, and they were carried out in the
framework  of the $t$-$J$ model which has the difficulty of the
constraint of no double occupancy  at the same site. In this paper we
return to the Hamiltonian $H^{(2)}_{p-d}$ derived in, {\em e.g.},
Refs.\cite{Zan,per,Shen,fulde}. This Hamiltonian is obtained from the
Emery model by means of a perturbative expansion up to the second
order in   $t_{pd}$ (the hopping parameter between $Cu$ and $O$
sites). We point out that the $t$ term of the $t$-$J$ Hamiltonian is
obtained from $H^{(2)}_{p-d}$ by projection on local singlet
states\cite{Zrice}. We will not consider here the  fourth order term
which gives the  super-exchange energy  between spins located on the
$Cu$ sites. This term will not be relevant in our calculation since we
want to study essentially the motion of a hole in a disordered
magnetic background.	In the framework of the $H^{(2)}_{p-d}$
Hamiltonian we will compute the dispersion relation $E(k)$  by
applying a trial wave function. The results agrees with previous
calculation for the $t$-$J$ model when we consider the limit of an
ordered  magnetic background.

 Our starting point is the $H^{(2)}_{p-d}$ Hamiltonian
 (for shortness we set  $H$ as  $H^{(2)}_{p-d}$ ) which can be
written as
\begin{equation}
H = \sum_{i,j,\sigma} T_{ij}\,f^\dagger_{i\sigma} f_{j\sigma}+g
\sum_{i,j,j',\sigma,\sigma'}  w_{ij}w_{ij'}\:
f^\dagger_{j\sigma} \:\vec{\tau}_{\sigma\sigma'} \:f_{j'\sigma'}.
\vec{S}_i
\label{p3.1}
\end{equation}
where $(f^+_{i\sigma},f_{i\sigma})$ is a fermion field with the spin
index $\sigma$. $\vec{S_i}$ is the spin operator on the $Cu$ site $i$,
$\vec{\tau}=(\tau_x,\tau_y,
\tau_z)$ are the Pauli matrices and \\
\begin{equation}
T_{ij} = T {1\over N} \: \sum_k \:\varepsilon_k\: {\mbox{e}}^{ik(i-j)}
\end{equation}
\begin{equation}
w_{ij} = {1\over N} \sum_k \:w_k
\: {\mbox{e}}^{ik(i-j)}
\end{equation}
where  $\varepsilon_k = 4(1-\gamma_k)$, $w_k = \sqrt{\varepsilon_k }$
and $\gamma_k = (\cos{k_x}+\cos{k_y})/2$.
The para\-me\-ters $T$ and $g$ are related to
the parameters $\Delta =\varepsilon_p - \varepsilon_d$, $U_d$
and $t_{pd}$ by the following relations\\
\begin{equation}
T \equiv {1\over 2}\;t^2_{pd} \: [{1\over\Delta} - {1\over{U_d -
\Delta}}]
\end{equation}
\begin {equation}
g \equiv t^2_{pd} \: [{1\over\Delta} + {1\over{U_d - \Delta}}]
\end{equation}\\
The Hamiltonian \rf{p3.1} describes a system of itinerant holes
interacting with a magnetic background. We notice that if we restrict
the $g$ term to a local spin-spin  interaction ({\em i.e.} take
$w_{ij} = w_0 \delta_{ij}$) we get the phenomenological Hamiltonian
considered by Monthoux and Pines\cite{pines93}.

\section{Construction of the  trial state for the one hole motion}

	The most general state for a single hole is
\begin{equation}
\psi = \sum_{i,\sigma}\: f^\dagger_{i\sigma} \: A(i\sigma) \: \Phi
\label{p3.2}
\end{equation}
where $\Phi$ is some magnetic background and $A(i\sigma)$ is some
combination of spin operators. The meaning of Eq.\rf{p3.2} is clear :
the motion of the hole will perturbe the  magnetic background
$\Phi$,   generating all sorts of  excited magnetic states provided by
the operator
$A(i\sigma)$. \newline
\noindent
In what follows, we will
suppose that the magnetic background $\Phi$
is  rotational invariant,{\em i.e.}
$$ {\vec S}_{tot} \Phi = 0$$\\
For the Hamiltonian under investigation Eq.\rf{p3.1}, if we set $g=0$,
the following
state for one hole with spin up
\beq
\sum_{i}\:\psi_o(i){f^\dagger_{i\uparrow}} \Phi
\label{p3.3}
\eeq
is an eigenstate for the $T$ term of the Hamiltonian Eq.\rf{p3.1},
if  $\psi_o(i)=\mbox{e}^{ik.i}$ for a given $k$.
On the other hand the $g$ term can be rewritten in
the following form
\beqa
\nonumber
& & -{3\over 4} g \sum_{i,j,j'}\: w_{ij}w_{ij'}\: [ \diuc \fjdc -
\didc\fjuc ]. [ \fjd \diu - \fju\did  ] +\\
& & {1\over 2}\; g \sum_{i,j,j'}\: w_{ij}w_{ij'}\:\left\{ \diuc\fjuc
\fju\diu +\didc\fjdc\fjd\did+\right.\nonumber\\
& & \left. {1\over 2}[ \diuc \fjdc +\didc\fjuc  ].
[ \fjd \diu + \fju\did  ]\right\}
\label{stform}
\eeqa
In this form we have splitted the Hamltonian into two  parts, one
which contains singlet operators and the  other one which contains
triplet operators. These singlet and triplet operators  are more
general than those considered by Zhang and Rice\cite{Zrice}, because
here they are non local.

\noindent
{}From Eq.\rf{stform} we see also that the  singlet  states and
the triplet states are separated by an energy $2g$. If we want
to look for low energy excitations and we neglect mixing between
the two bands, we can write the following state for the
one hole with spin up
\beq
\sum_{i,j}\:\psi_1(i,j)\: \left[ \diuc \fjdc - \didc\fjuc \right
]\did\Phi
\label{naturel}
\eeq
Then a natural trial state for the full Hamiltonian Eq.\rf{p3.1}
can be written as
\beq
\sum_{i}\:\psi_o(i){f^\dagger_{i\uparrow}} \Phi+
\sum_{i,j}\:\psi_1(i,j)\: \left[ \diuc \fjdc - \didc\fjuc \right
]\did\Phi
\label{state}
\eeq
To simplify calculation we take $\psi_1(i,j)=\varphi_1(j) w_{ji}$.
Then the state \rf{state} can be rewritten in a more
transparent way

\begin{equation}
\psi = \sum_i\left\{ \varphi_o(i){f^\dagger}_{i\uparrow} +
\sum_j \varphi_1 (j) w_{ij}
\left [f^\dagger_{i\uparrow} {S^z\!}_j + {f^\dagger}_{i\downarrow}
S^{+}_j\right ]\right\} \Phi
\label{p3.4}
\end{equation}
where $\varphi_o(i)=\psi_o(i)-{1\over 2}\sum_j \varphi_1 (j)
w_{ij}$.\\

 Examples  for the magnetic state $\Phi$ are the singlet
Resonating-Valence-Bond states  considered by Anderson, Dou\c cot and
Liang\cite{liang} which may be long ranged or short ranged ordered.

\section{ Dispersion relation for a single hole}

	The dispertion relation $E(k)$ will be calculated by using the
variational method. We will minimize $(\psi, H\psi)$  keeping the norm
$(\psi, \psi)$ fixed. Straightforward calculation, using the property
: $\sum_l\;w_{il}w_{lj} = {\hat T}_{ij}$, \ leads to

$$(\psi, H\psi) =\sum_{i,j}\; T_{ij}\;\left\{ \overline{\varphi_o(i)}
\varphi_o(j) +
\sum_{l,l'} \;w_{il} w_{jl'}\; \overline{\varphi_1(l')}
\varphi_1(l) <\vec{S_{l'}}\vec{S_l} > \right\}+$$

$$+g \sum_{j,l,l'} \;{\hat T}_{ll'} w_{jl} \;
\left\{ \overline{\varphi_1(l')} \varphi_o(j) < \vec{S_{l'}}\vec{S_l}
> + h.c\right\} -$$
$$-g \sum_{l,l'}\; {\hat T}_{ll'}\;\left\{
\overline{\varphi_1(l)} \varphi_1(l') < \vec{S_l}\vec{S_{l'}} > +
h.c\right\}+$$

$$+ g \sum_{l,l'}\; {\hat T}_{ll'}{\hat T}_{ll'} \;
\overline{\varphi_1(l')} \varphi_1(l)  < \vec{S_l} \vec{S_{l'}} >$$
\beq
+ ig\sum_{{l_1,l_2,l_3}\; (l_1\neq l_2 \neq l_3)}
{\hat T}_{l_1 l_3} {\hat T}_{l_1 l_2} \; \overline{\varphi_1(l_3)}
\varphi_1(l_2)\;
 <\vec{S_{l_3}} . (\vec{S_{l_1}} \times \vec{S_{l_2}}) >
\label{p3.5}
\eeq
with ${\hat T}_{ij}=T_{ij}/T$ and  the average $<...>$ defined as
\beq
<S^a ...S^b>\equiv (\Phi,S^a ...S^b\Phi)
\eeq

Now if we suppose that the magnetic background $\Phi$ [in
Eq.\rf{p3.4}] is invariant under the time reversal operation $K$ ,
 we will have
\beqa
\nonumber
\left( K\Phi, {S^x}_{l_1}{S^y}_{l_2}{S^z}_{l_3} K\Phi\right) & = &
\left( \Phi, {S^x}_{l_1}{S^y}_{l_2}{S^z}_{l_3} \Phi\right)\\
&=& -\left( \Phi, {S^x}_{l_1}{S^y}_{l_2}{S^z}_{l_3} \Phi\right)
\eeqa
for any $l_1$, $l_2$, $l_3$ (all different), since $K {\vec S}
K^{-1}  = -\vec{S}$. And finally we have
\beq
<\vec{S_{l_3}} . (\vec{S_{l_1}} \times \vec{S_{l_2}}) >= 0
\eeq
Consequently the last term of Eq.\rf{p3.5} vanishes.\\
The energy will be calculated by minimizing the energy functional
\beq
{\cal W}(\psi , \lambda) = (\psi, H\psi) -\lambda (\psi,\psi)
\eeq
\noindent
Using Eq.\rf{p3.5} and

\beq
( \psi ,\psi ) = \sum_i {\mid \varphi_o(i)\mid}^2 +
\sum_{j,j'} {\hat T}_{jj'} {\overline{\varphi_1(j)}} \varphi_1(j')
<\vec{S_j} \vec{S_{j'}}>
\label{p3.6'}
\eeq
we get the following equation for the energy $E$
\\
\beq
\left(
\ba{cc}
T \varepsilon_k - E  &  g w_k a_k \\
\\
g w_k a_k      &  T c_k + g d_k - E a_k
\ea
\right)
{}.\left(
\ba{l}
\varphi_o(k) \\
\\
\varphi_1(k)
\ea
\right) ={\bf 0}
\label{p3.6}
\eeq
\\
\noindent
where $\varphi_o(k)$ and $\varphi_1(k)$ are the Fourier components of
$\varphi_o(i)$ and
$\varphi_1(i)$ respectively.
On the other hand
\\
\beq
\left\{
\ba{l l}
a_k  =  3 - 4 \chi_{o1}\; \gamma_k \; ; &  c_k  = 15-32\chi_{o1}\;
\gamma_k +
4\chi_{o2}(4\gamma_k^2-1)+4\gamma_{2k}(\chi_{11}-\chi_{o2})\\
\\
 d_k  = 12+\chi_{o1}-8 a_k &
\ea
\right.
\label{p3.7'}
\eeq
\\
with
\beq
\chi_{o1} = -<{\vec S}_o {\vec S}_{e_1} > \; ; \;
\chi_{11} = <{\vec S}_o {\vec S}_{e_1 + e_2} > \; \; \mbox{and} \; \;
\chi_{o2} = -<{\vec S}_o {\vec S}_{2e_1}>
\eeq
$e_1$ and $e_2$ denoting the two unit vectors of the square lattice.
{}From  Eq.\rf{p3.6},   the  lowest  energy band   is given by
\beqa
\nonumber
E_k/g & = &{1\over 2}\left\{ \eta (\varepsilon_k + {{c_k}
\over {a_k}}) + {{d_k} \over{a_k}} - \right.\\
 & - & \left. \sqrt {[\eta (\varepsilon_k - {{c_k} \over {a_k}}) -
{{d_k} \over {a_k}}]^2 + 4\varepsilon_k a_k}\right\}
\label{energy}
\eeqa
with $\eta = T/g$ and for a given $k$ we have the following solutions
for
$\varphi_o(i)$ and $\varphi_1(i)$
\beqa
\nonumber
\varphi_o(i;k) & = &  w_k a_k U(k){\mbox{e}}^{i k.i}\\
\varphi_1(i;k) & = & - (\eta\varepsilon_k - E_k/g) U(k){\mbox{e}}^{i
k.i}
\label{p3.9}
\eeqa
where the function $U(k)$ is determined by writing $(\psi, \psi) = 1$
and we get
\beq
U(k)^2 = {1\over { \varepsilon_k a^2_k + {3\over 4} [\eta
\varepsilon_k - E_k/g]^2}}
\label{p3.9'}
\eeq
 The
dispersion relation $E_k$ depends  only on the the spin-spin
correlation function for nearest and next nearest neighbors.
This is due to our expression for the trial state. Introducing more
and more
spin excitations would essentially produce new spin
correlation functions at larger distance. Hower since
we are interested essentially in a background with short correlation
functions, this would merely not chane our result at least
qualitatively.
 Knowing
$\chi_{o1}$, $\chi_{11}$ and $\chi_{o2}$  we evaluate then the energy
$E_k$.

If we set  $\eta =0$, we can have a qualitative idea about the
positon of the minimum of $E_k$
 \beq
E_k/g = -{1\over 2}\left\{ \mid{{d_k} \over{a_k}}\mid +
  \sqrt {\mid{{d_k} \over {a_k}}\mid^2 +
4\varepsilon_k a_k}\right\}
\eeq
It is easy to see that $\mid{{d_k}/{a_k}}\mid$ is maximal for
$k=(0,0)$ and $4 \varepsilon_k a_k$ is maximal for $k=(\pi,\pi)$. Then
since to lower $E_k$ both of the quantities should be maximal, there
will be a competition between them. For  $\chi_{o1}=0$ ,
$\mid{{d_k}/{a_k}}\mid = 4$ and consequently the minimum will be at
$k=(\pi,\pi)$. For $\chi_{o1}\neq 0$ the position of the minimum will
be in  between $(0,0)$ and $(\pi,\pi)$.

\subsection{ Paramagnetic case}

For a completely disordered magnetic system, we have
\beq
<\vec{S_i} \vec{S_j}> = {3\over 4} \delta_{ij}
\label{p3.7}
\eeq
Therefore $\chi_{o1}$, $\chi_{11}$  and $\chi_{o2}$ vanish
and we obtain for the energy

\beq
E_k/g = \eta ({9\over 2} -2\gamma_k) - 2- \sqrt{12(1-\gamma_k) + [2 -
\eta  ({1\over 2}+ 2\gamma_k)]^2}
\label{p3.8}
\eeq
The dispersion relation Eq.\rf{p3.8} is plotted in Fig.\rf{disord}
for the  physical value  $\eta=1/6$. The band minimum is reached at
points $(\pm\pi, \pm\pi)$. The center of the Brillonin zone
$\Gamma(0,0)$ is a maximum. The bandwidth is given by

\beqa
\nonumber
W & = & E(0,0) - E(\pi,\pi) \\
  & = & 2g \left\{ \sqrt{6 + [ 1 + {3 \over 4} \eta]^2} - 1 - {3\over
4}\eta\right\}
\eeqa

\noindent
for the physical value  $\eta =1/6$, we have $  W= 3.14g$. From
Ref.\cite{Shen} we can get the relation between $g$ and the $t$
parameter of the $t$-$J$ model, and we have $t=g(3-2\eta)/4$. We end
then with a bandwith $W\simeq 4.71 t$. We obtain  a bandwidth of the
order of $t$. We notice that even in this disordered case the
bandwidth was reduced from $8t$ (for the $t$-$J$ model  with $J = 0$ :
the free motion) to $4.71t$, which express the strong correlation in
the system. We expect that in the ordered state the bandwidth will be
reduced  further.

Now we want to examine the spin pattern around the hole. We  started
{}from  a state without order and we want to calculate the local
magnetization at site $j$, if the hole is at site $i$ for the trial
wave function $\psi^{(k)}$ with energy $E_k$.
Algebric calculation leads to

\beq
\frac {(\psi^{(k)}, n_i \szj \psi^{(k)})} {(\psi^{(k)}, \psi^{(k)})}
=  - w_{ij}\;
\cos k(i-j)\;
\frac
{w_k a_k (\eta \varepsilon_k - E_k/g)}
{\varepsilon_k a^2_k + 3 [\eta \varepsilon_k - E_k/g]^2}
\eeq
where we have used Eqs.\rf{p3.9}\rf{p3.9'}.
Since the minimum of energy is at $Q = (\pi, \pi)$, the hole will
have the lowest energy $E_Q$ and then  we get the
following correlation function between the hole and surrounding spins
\beq
<n_i \szj> \equiv \frac {(\psi^{(k)}, n_i \szj \psi^{(k)})}
{(\psi^{(k)}, \psi^{(k)})}
 = {1\over {\sqrt 2}}
\left[ \frac { E_Q/g - 8\eta} {6 + {1\over 4}[E_Q/g-8\eta]^2  }
\right] \; w_{ij}\;
 \cos Q(i-j)
\eeq
For  $\eta = 1/6$ we obtain

\beq
<n_i \szj>\simeq -{1\over 4}\; w_{ij}\; \cos Q(i-j)
\label{p3.10}
\eeq
 Eq.\rf{p3.10} means that the hole polarizes the spins in its
vicinity. The spin pattern around the hole is antiferromagnetic and
the correlation decreases like
$w_{ij}$ when the distance between the hole and neighbor spin
increases (see Fig.\rf{spins}).

The trial state $\psi$  corresponds
to a state with a total spin  $1/2$ and $S^z_{tot} = + {1/2}$.
The state corresponding to $S^z_{tot} = - {1/ 2}$ is just given by
 $S^-_{tot}\psi$. And the correlation function  for a hole with spin
down is related to the case of spin up by

\beqa
\nonumber
<n_i \szj>^{(\downarrow)} & = & (S^-_{tot} \psi, n_i \szj
{S^{-}}_{tot} \psi) \\
                           & = & - <n_i \szj>^{(\uparrow)}
\eeqa

This implies  that a hole with spin down will polarize spins in its
vicinity with opposite direction  with respect to a hole with  spin
up. Naively we may say that when two holes with opposite spins are
introduced in the system  it will be preferable to have them close to
each other in order to cancel their spin polarizations and obtain
again a system without magnetic order which favours the motion of the
holes. This scenario might be similar to the  Cooper pairing.

\subsection{ Locally ordered case}

Since experiments on copper oxides show that the spin susceptibility
presents a sharp peak near the wave vector $Q$\cite{RossRev}
in the metallic phase, we shall take
a lorentzian shape
for the static susceptibility to describe the case of a locally
ordered state. We write then

\beq
\chi_q = \frac {C(\xi)} {1 + \xi^2 [1 + {1\over 2} (\cos q_x + \cos
q_y)]}
\label{p3.11}
\eeq

which may be correct only for short correlation lenght  $\xi$ .
$C(\xi)$ is determined by writing $<\vec{S_i} \vec{S_i}> = {3/4}$
which gives

\beq
C(\xi) = {3\over 4}\left[{{1\over N} \sum_q {1\over {1 + \xi^2 [1
+{1\over 2}  (\cos q_x + \cos q_y)]}}}\right]^{-1}
\eeq
 The inverse Fourier
Transform of Eq.\rf{p3.11} gives
\beqa
\nonumber
\chi(i-j) & = & {1\over N} \sum_q \frac { C(\xi){\mbox{e}}^{iq(i-j)}}
{1 + \xi^2 [1 + {1\over 2} (\cos q_x + \cos q_y)]}\\
\nonumber
\\
 & = & \frac {4 C(\xi)} {\xi^2}  \chi_Q (i-j){\mbox{e}}^{-iQ(i-j)}
\eeqa
and $\chi_Q(i-j)$ satisfies the equation

\beq
\sum_l\;(-\Delta)_{il}\; \chi_Q (l-j)+ {4 \over {\xi^2}} \chi_Q (i-j)
= \delta_{ij}
\eeq
with $\Delta$ the Laplacian on the square lattice. The asymptotic
solution of this equation ( i.e. for $\mid i-j \mid \rightarrow
\infty$)
 is known\cite{McB} and we have

\beq
\chi_Q (i-j) \sim {\mbox{e}}^{- \mid i-j \mid / \xi_L}
\eeq

where $\xi_L$ is given  by

\beq
{1\over {\xi_L}} = \mbox{ch}^{-1} (1 + {1\over \xi^2})
\eeq

For $\xi$ sufficiently large  we have
$ \xi_L \simeq \xi/\sqrt{2}$.

	Now having the phenomenological form of the static susceptibility
(Eq.\rf{p3.11}),  we can compute $\chi_{o1}$, $\chi_{11}$ and
$\chi_{o2}$ for different correlation length. The table (1) contains
some values of these quantities. We notice that by increasing $\xi$ we
can have more and more ordered state. For example we almost recover
the values of $\chi_{o1}$, $\chi_{11}$ and $\chi_{o2}$ predicted by
the linear spin wave approximation for $\xi=5$.
\vskip 0.3truecm
\begin{center}
\begin{tabular}{llll}
\hline
\hline
\\
$  \xi\   $ &        $ \chi_{o1}  $     &
$ \chi_{11}  $     &  $\chi_{o2} $     \\
\\
\hline
\\
1   &          -0.105        &        0.0278          &     0.015        \\
 2   &          -0.199        &        0.094          &     0.058        \\
 3   &          -0.259        &        0.150          &     0.101        \\
 4   &          -0.299        &        0.192          &     0.138        \\
 5   &          -0.328        &        0.223          &     0.167        \\
 \\
\hline
\hline
\end{tabular}
\end{center}
\begin{center}
\begin{minipage}{10cm}
{\small {\bf Table 1 :} Correlation functions  $\chi_{o1}$,
$\chi_{11}$ and $\chi_{o2}$ calculated from the phenomenological form
of the static susceptibility Eq.\rf{p3.11}, for different $\xi$. }
\end{minipage}
\end{center}
\vskip 0.5truecm

Inserting the  values of $\chi_{o1}$, $\chi_{11}$ and $\chi_{o2}$
 in Eq.\rf{energy} we get the dispersion plotted in Fig.\rf{sw}
for different values of $\xi$. We see from Fig.\rf{sw} that the
spectrum changes drastically when the magnetic background changes from
a disordered state to an ordered one.
The minimum of $E_k$
shifts from $(\pm \pi, \pm \pi)$ in the case of a disordered
background $(\xi = 0)$
 to $(\pm \pi/2, \pm \pi/2)$ for $\xi=3$. We see clearly that the
bandwidth has been strongly reduced in comparison with the disordered
case.  It was reduced from $W\simeq 3.14g$ ( for $\xi=0$)
to $W\simeq 1.3g $ ( for $\xi=3$). In term of $t$ we get a bandwidth
$W\simeq 1.8t$.
 But it is still of the order of $t$ and not $J$. This is
a discrepancy with previous calculations done in the framework of the
$t$-$J$ model for an AFM background\cite{Varma,Kane,MHorsch,JLR} which
may be explained by the fact that no magnetization is
present in our calculation since we considered our background as a
singlet state. Other  reason  might be that with our
variational approach we do not include the incoherent motion of the
quasi-particle which is the consequence of retarded effects.

\section{Conclusion}
	In this section we have computed the dispersion relation $E(k)$ for
a single hole in the framework of the spin-hole model using a
variational method. The calculation was done for any rotational
invariant magnetic background. And we have shown how the hole spectrum
changes when the magnetic environment goes from  a disordered
background to an ordered one. Especially we have found that the
minimum energy  shifts from $(\pm \pi, \pm \pi)$ for the completely
disordered case to $(\pm\pi /2, \pm\pi /2)$ in the case of ordered
magnetic background. The bandwidth is strongly reduced by a factor
1/3, when the magnetic background changes from a disordered to an
ordered state.
 An other interesting result that we
 have found is the spin polarization around the hole. Namely when a
hole moves in a completely disordered backgound it will polarize spins
in its vicinity. This polarization is decreasing with  the distance as
a power law and has an AFM ordering.
 We expect that two holes with opposite spins will tend to
pair in order to remove their respective spin polarizations. This
question  deserves further studies and it is now under investigation.

\eject

\newpage
\begin{figure}
\vbox{\vskip 5truecm
\hskip 1.5truecm \special{illustration Figure1 scaled 500}
}
\vskip -5truecm
\begin{center}
\begin{minipage}{10cm}
\caption{\em Dispersion relation for a single hole in disordered
magnetic background( with $M=(\pi,\pi)$ , $\Gamma=(0,0)$ and $X=
(\pi,0)$). $E_k$  is in unit of $g$.}
\label{disord}
\end{minipage}
\end{center}
\end{figure}

\begin{figure}
\vbox{\vskip 7truecm
\hskip 3.5truecm \special{illustration Figure2 scaled 340}
}
\begin{center}
\begin{minipage}{10cm}
\caption{\em Polarization of spins surrounding the hole.}
\label{spins}
\end{minipage}
\end{center}
\end{figure}

\begin{figure}[t]
\vbox{\vskip 7truecm
\hskip -1truecm \special{illustration Figure3 scaled 900}
}
\vskip -5truecm
\begin{center}
\begin{minipage}{10cm}
\caption{\em Dispersion relation for a single hole in different
magnetic backgrounds : from a  completey disordered state ($\xi =0$)
to an ordered state $\xi=5$. $E_k$ is in unit of $g$ ( with
$M=(\pi,\pi)$ , $\Gamma=(0,0)$ and $X=(\pi,0)$).}
\label{sw}
\end{minipage}
\end{center}
\end{figure}

\end{document}